\documentclass{ieeeaccess}
\usepackage{cite}
\usepackage{amsmath,amssymb,amsfonts}
\usepackage{algorithmic}
\usepackage{graphicx}
\usepackage{textcomp}
\usepackage{float}
\usepackage{multirow}
\usepackage{multicol}
\usepackage{booktabs}
\usepackage{longtable}
\usepackage{subcaption}

\usepackage{caption,setspace}
\def\BibTeX{{\rm B\kern-.05em{\sc i\kern-.025em b}\kern-.08em
    T\kern-.1667em\lower.7ex\hbox{E}\kern-.125emX}}
\begin{document}
\history{}
\doi{}

\title{Trajectory Tracking Control of UAV Bicopter using Linear Quadratic Gaussian}
\author{\uppercase{Fahmizal}\authorrefmark{1,$\circ$,$\lozenge$}, 
\uppercase{Hanung Adi Nugroho}\authorrefmark{2,$\circ$}, 
\uppercase{Adha Imam Cahyadi}\authorrefmark{3,$\circ$},
and \uppercase{Igi Ardiyanto}\authorrefmark{4,$\circ$}}

\address[$\circ$]{Department of Electrical and Information Engineering, Engineering Faculty, Universitas Gadjah Mada, 55281, Indonesia}
\address[$\lozenge$]{Department of Electrical Engineering and Informatics, Vocational College, Universitas Gadjah Mada, 55281, Indonesia}

\markboth
{Fahmizal \headeretal: Trajectory Tracking Control of UAV Bicopter using Linear Quadratic Gaussian}
{Fahmizal \headeretal: Trajectory Tracking Control of UAV Bicopter using Linear Quadratic Gaussian}

\corresp{Email: $^{1}$fahmizal@ugm.ac.id, $^{2}$adinugroho@ugm.ac.id, $^{3}$adha.imam@ugm.ac.id, $^{4}$igi@ugm.ac.id}

\begin{abstract}
This paper describes the design of a linear quadratic gaussian (LQG) for trajectory tracking control of UAV Bicopter. In this work, disturbance in the form of payload significantly affects the trajectory tracking control process on the UAV Bicopter when using a linear quadratic regulator (LQR) controller. The use of a LQR control will be optimal in the case of a state regulator towards an equilibrium point in a system, but for the tracking case, the LQR controller is not capable of optimally, especially in systems that have high levels of nonlinearity and system dynamic changes such as inertial disturbances. Therefore, this paper proposes the design of a LQG control that is expected to overcome system dynamic changes, in this case in the form of inertial disturbances to the UAV Bicopter when carrying a payload. The success of LQG control was tested in two scenarios, the first trajectory tracking at a circular position and the second with the position of the trajectory number ”8”. The simulation results show that the proposed LQG controller successfully overcame inertial disturbances when the UAV Bicopter performs trajectory tracking. When given an inertial disturbance, the trajectory tracking test results show that the LQG control has a lower root mean square error (RMSE) value than the LQR control.
\end{abstract}


\titlepgskip=-21pt

\maketitle

\section{Introduction}

\normalfont Over the last few decades, the use and development of Unmanned Aerial Vehicles (UAV) has expanded, ranging from hobbyists to military applications. Several other applications apply to surveying \cite{cantelli2013uav,turner2016uavs, bonali2019uav}, maintenance and surveillance tasks \cite{kannadaguli2020yolo, dwivedi2018maraal,azmat2023elliptical}, transport and manipulation \cite{menouar2017uav,gupta2021advances}, search and rescue \cite{alotaibi2019lsar,ahn2018reliable,mcconkey2023semi}. This type of UAV using a propeller is very attractive because it offers advanced capabilities such as Vertical Take Off and Landing (VTOL) and high agility. Quadcopters-type UAVs are the most researched and used platforms in this regard.

However, the UAV Quadcopter has limits when applied to indoor environments. Indoor UAV applications usually demand a high payload capacity at a small size while having sufficient endurance to complete tasks/missions. This requirement has caused the ability of Quadcopters not to fulfill it. For example, the DJI Matrice M100 \cite{matrice100} can carry up to 0.9 kg of weight but has a wingspan of 806 mm, which is far greater than the average door width in indoor operation. An intuitive approach is to reduce the size of the UAV so that it can fit through the door frame. However, there are requirements for UAVs to be able to carry quite a heavy payload, so there are other solutions than reducing size. As another example, Crazyflie 2.0 is a mini-UAV Quadcopter capable of crossing very small gaps \cite{giernacki2017crazyflie}. However, its maximum payload mass is only 15 g.  

Adaptive morphology is another solution to allow UAVs to adapt to confined spaces \cite{de2007artificial,mintchev2016adaptive,bucki2019design,derrouaoui2021towards,martynov2023morphogear}. Research conducted by D. Falanga et al. \cite{falanga2018foldable} changed the UAV frame into a different configuration to let it go through areas it might not have been able to. This solution allows the UAV to maintain size without sacrificing power and efficiency. However, the overall deformation rate and size of the UAV are constrained by the large number of propellers used. In addition, four additional servomotors also increase the mechanical complexity and weight of the UAV.

Even to reduce the size of the UAV, Riviere et al. \cite{riviere2018agile} proposed a morphological mechanism that arranges all four rotors in succession, reducing the size of the UAV to the size of a single propeller. However, the rolling motion could not be controlled with such a transformation. This paper offers a UAV design solution using a tilt-rotor mechanism by selecting the Bicopter UAV type. The advantage of this type of UAV Bicopter is that it can be used for flying exploration in indoor applications or narrow spaces, such as passing through the size of a door or window in a house. In addition, the number of sets of actuators is less when compared to UAV Quadcopters or multi-rotor types. As a result of the size, the required power consumption is also more efficient. Besides its superiority in efficiency and power size, the UAV Bicopter is very challenging to control, especially in relation to altitude and attitude control on the UAV Bicopter, especially in the case of trajectory tracking.

The controller design for the attitude of the UAV Bicopter is important because attitude control is a mechanism for controlling the orientation of the UAV Bicopter in the form of roll, pitch and yaw rotational movements. As an inner loop in the Bicopter UAV system, the rotational motion system must have a fast settling time to support the translational motion system as an outer loop or position control. The problem that arises in this position control is tracking. The translational motion system must be able to follow the given reference signal and overcome the given disturbance.

This paper contributes to formulating a UAV Bicopter mathematical model by bringing it closer to a linear system and simulating flight control of a UAV Bicopter in the case of trajectory tracking using an optimal linear quadratic regulator (LQR) controller when there is no inertial disturbance, but when there is an inertial disturbance using an optimal linear quadratic controller Gaussian (LQG).

This paper's remaining sections are organized as follows: Section II covers the dynamics of Bicopter and provides inertial disturbance modeling. Section III describes the design of the LQG controller for trajectory tracking control. Simulation results are presented in Section IV to demonstrate the value and effectiveness of the proposed methodologies. Section V concludes the paper.

\section{Methodology}
\subsection{Dynamics Modelling of a Bicopter}
This sub-chapter discusses the modeling of the Bicopter system observed in a Cartesian diagram frame with three-dimensional axes ($x$, $y$, $z$). The frame diagram is divided into two, namely the earth frame which is not moving, and the Bicopter body frame (body frame). The linear position of the Bicopter ($\Gamma ^{E}$) is determined from the vector coordinates between the origin of the body frame (B-frame) and the origin of the earth frame (E-frame) with respect to the Eframe. The angular position of the Bicopter ($\Theta ^{E}$) is determined from the orientation of the B-frame to the E-frame. The linear position and angular position are found in Eq. (\ref{eq3w30}) - (\ref{eq3w31}). Next, the Bicopter rotates about the ($x$, $y$, $z$) axis using the rotation matrix.

\begin{equation}
	\Gamma ^{E}=\left [ \begin{matrix}
		x  &y   &z  
	\end{matrix} \right ]
	\label{eq3w30}
\end{equation}
\begin{equation}
	\Theta ^{E}=\left [ \begin{matrix}
		\phi  &\theta   &\psi  
	\end{matrix} \right ]
	\label{eq3w31}
\end{equation}

\begin{figure}[h]
	\centering
	\includegraphics[scale=0.24]{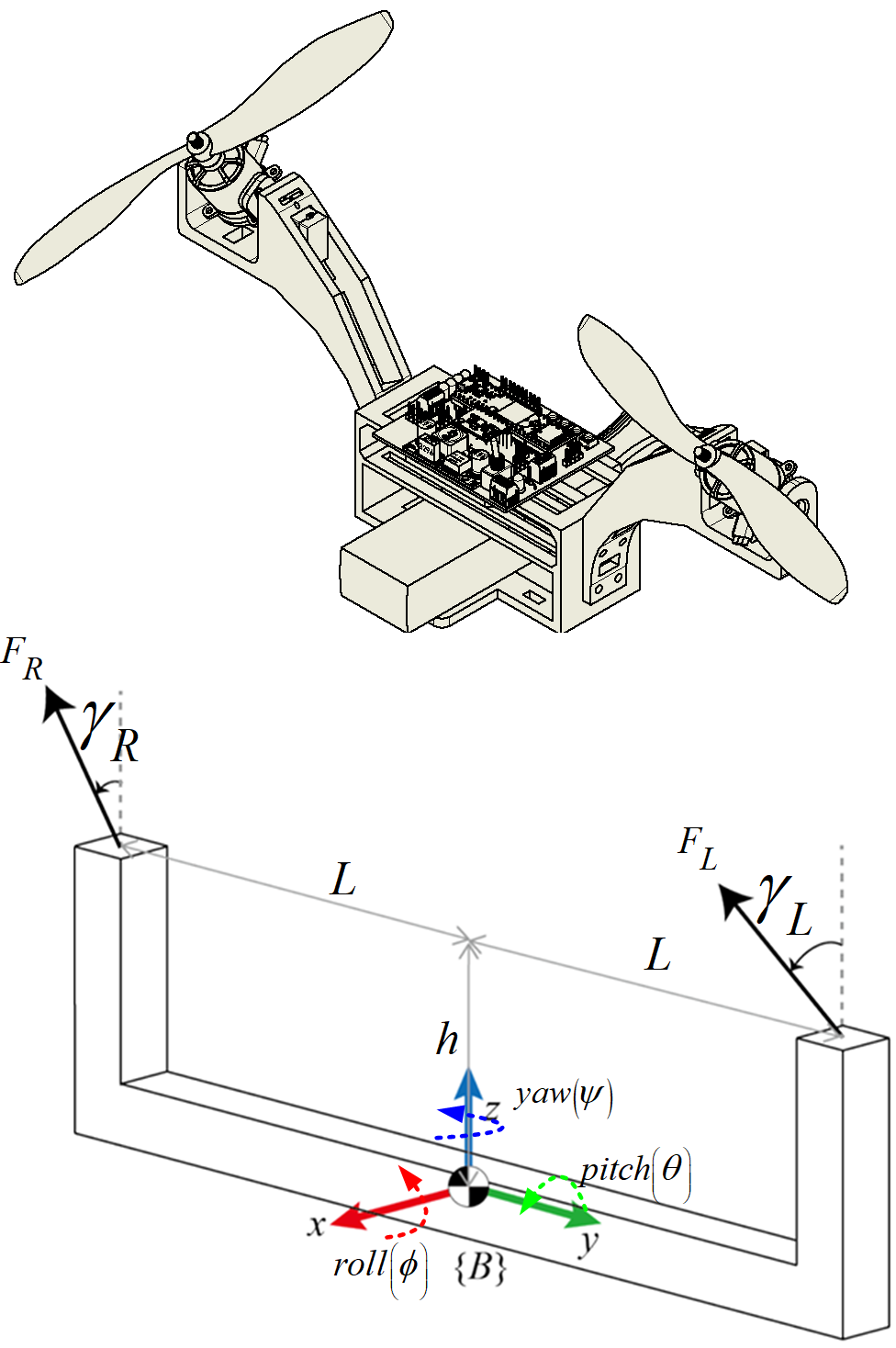}
	\caption{Body diagram of Bicopter.}
	\label{fig:gb3w14}
\end{figure}

Bicopter speed consists of linear velocity $\left [ \begin{matrix}\dot{x}  &\dot{y}   &\dot{z} \end{matrix} \right ]$ and angular velocity $\left [ \begin{matrix}\dot{\phi }  &\dot{\theta }   &\dot{\psi } \end{matrix} \right ]$ so that when represented in state space form, the Bicopter equation has twelve states which are divided into two categories, six states translation and six rotational states as in Eq. (\ref{eq3w32}) - (\ref{eq3w33}). Right rotor thrust ($F_{R}$) and left rotor thrust ($F_{L}$) generated by the propeller and rotor and their components in the $x$ and $z$ directions are shown in Fig. \ref{fig:gb3w14} and with the parameters described in Table \ref{tab:parameterBicopter}. In addition, the right side tilt angle and left are denoted as $\gamma  _{R}$ and $\gamma  _{L}$. Using Newton's second law, the equations of forces in the $x$, $y$ and $z$ directions are defined as given in Eq. (\ref{eq3w34}) - (\ref{eq3w36}).

\begin{equation}
	x_{1-6}=\left [ \begin{matrix}x  &\dot{x}   &y &\dot{y} &z &\dot{z}\end{matrix} \right ]^{T}
	\label{eq3w32}
\end{equation}
\begin{equation}
	x_{7-12}=\left [ \begin{matrix}\phi  &\dot{\phi }   &\theta &\dot{\theta } &\psi &\dot{\psi }\end{matrix} \right ]^{T}
	\label{eq3w33}
\end{equation}




\begin{equation}
	\sum F_{x}=F_{R}\sin \gamma  _{R}+F_{L}\sin \gamma  _{L}
	\label{eq3w34}
\end{equation}
\begin{equation}
	\sum F_{y}=0
	\label{eq3w35}
\end{equation}
\begin{equation}
	\sum F_{z}=F_{R}\cos \gamma  _{R}+F_{L}\cos \gamma  _{L}
	\label{eq3w36}
\end{equation}

From the input $u$ in Eq. (\ref{eq3w37}) – (\ref{eq3w41}), we can calculate the Bicopter's total lift (thrust) and moment of force. Where $C_{T}$ is the propeller's thrust coefficient. $\Omega _{R}$ and $\Omega _{L}$ represent the rotors' right and left rotational speeds, and $\gamma  _{R}$ and $\gamma  _{L}$ represent the rotors' right and left tilt angles.

\begin{equation}
	u=\left [ \begin{matrix}
		u_{1} & u_{2} & u_{3} & u_{4}
	\end{matrix} \right ]^{T}
	\label{eq3w37}
\end{equation}
\begin{equation}
	u_{1}=C_{T}\left ( \Omega _{R}^{2}\cos \gamma  _{R} +\Omega _{L}^{2}\cos\gamma  _{L}\right )
	\label{eq3w38}
\end{equation}
\begin{equation}
	u_{2}=C_{T}\left ( \Omega _{R}^{2}\cos \gamma  _{R} -\Omega _{L}^{2}\cos\gamma  _{L}\right )
	\label{eq3w39}
\end{equation}
\begin{equation}
	u_{3}=C_{T}\left ( \Omega _{R}^{2}\sin \gamma  _{R} +\Omega _{L}^{2}\sin\gamma  _{L}\right )
	\label{eq3w40}
\end{equation}
\begin{equation}
	u_{4}=C_{T}\left ( \Omega _{R}^{2}\sin \gamma  _{R} -\Omega _{L}^{2}\sin\gamma  _{L}\right )
	\label{eq3w41}
\end{equation}

The rotation subsystem (roll, pitch, and yaw) is the inner loop, and the translation subsystem ($x$, $y$ position, and $z$ (altitude)) is the outer loop. Based on the dynamic solution of the model using Newton-Euler  \cite{albayrak2019design,zhang2016modeling, abedini2021robust}, we get Eq. (\ref{eq3w42}) for translational motion and Eq. (\ref{eq3w43}) for circular motion, where $s = sin$ and $c = cos$.

\begin{align}
	&\ddot{x}=-\frac{1}{m}\left ( s\phi s\psi +c\phi s\theta c\psi  \right )u_{1}-\frac{c\theta c\psi }{m}u_{3} \nonumber \\
	&\ddot{y}=-\frac{1}{m}\left ( -s\phi c\psi +c\phi s\theta s\psi  \right )u_{1}+\frac{c\theta s\psi }{m}u_{3} \nonumber \\
	&\ddot{z}=g-\frac{1}{m}\left ( c\phi c\theta  \right )u_{1}-\frac{s\theta }{m}u_{3}
	\label{eq3w42}
\end{align}

\begin{align}
	&\ddot{\phi }=\frac{L}{I_{xx}}u_{2} \nonumber \\
	&\ddot{\theta }=\frac{h}{I_{yy}}u_{3} \nonumber \\
	&\ddot{\psi }=\frac{L}{I_{zz}}u_{4}
	\label{eq3w43}
\end{align}

Equation (\ref{eq3w44}) represents the Bicopter's state space if reorganized based on Eq. (\ref{eq3w32}) – (\ref{eq3w33}).

\begin{align}
	&\dot{x}_{1}=\dot{x}=x_{2} \nonumber \\
	&\dot{x}_{2}=\ddot{x}=-\frac{1}{m}\left ( s\phi s\psi +c\phi s\theta c\psi  \right )u_{1}-\frac{c\theta c\psi }{m}u_{3} \nonumber \\
	&\dot{x}_{3}=\dot{y}=x_{4} \nonumber \\
	&\dot{x}_{4}=\ddot{y}=-\frac{1}{m}\left ( -s\phi c\psi +c\phi s\theta s\psi  \right )u_{1}+\frac{c\theta s\psi }{m}u_{3} \nonumber \\
	&\dot{x}_{5}=\dot{z}=x_{6} \nonumber \\
	&\dot{x}_{6}=\ddot{z}=g-\frac{1}{m}\left ( c\phi c\theta  \right )u_{1}-\frac{s\theta }{m}u_{3} \nonumber \\
	&\dot{x}_{7}=\dot{\phi}=x_{8} \nonumber \\
	&\dot{x}_{8}=\ddot{\phi }=\frac{L}{I_{xx}}u_{2} \nonumber \\
	&\dot{x}_{9}=\dot{\theta}=x_{10} \nonumber \\
	&\dot{x}_{10}=\ddot{\theta }=\frac{h}{I_{yy}}u_{3} \nonumber \\
	&\dot{x}_{11}=\dot{\psi}=x_{12} \nonumber \\
	&\dot{x}_{12}=\ddot{\psi }=\frac{L}{I_{zz}}u_{4}
	\label{eq3w44}
\end{align}

In order to develop a linear quadratic regulator (LQR) controller, it is necessary to linearize the nonlinear model described in Eq. (\ref{eq3w44}) around a specific operating point. Typically, this operating point is selected to be in hovering conditions. The assumption is made that the UAV Bicopter's hovering behavior may be described by Eq. (\ref{hovering}).

\begin{equation}
	\begin{split}
		u_{1}\simeq mg\\
		u_{3}\simeq 0\\
		sin\left ( \phi  \right )\simeq \phi\rightarrow \phi \simeq 0,cos\left ( \phi \right )\simeq 1\\
		sin\left ( \theta  \right )\simeq \theta\rightarrow \theta \simeq 0,cos\left ( \theta \right )\simeq 1\\
		sin\left ( \psi  \right )\simeq 0\rightarrow cos\left ( \psi \right )=\sqrt{1-sin^{2}\left ( \psi \right )}=1 
	\end{split}
	\label{hovering}
\end{equation}

During this hovering condition, the attitude yaw angle of the Bicopter ($\psi$) is assumed to be zero. Under these conditions, the dynamics of the attitude pitch ($\theta$) will be responsible for controlling the $x$ position of the UAV Bicopter, and the dynamics of the attitude roll ($\phi$) will control the $y$ position of the UAV Bicopter. By using the assumption of hovering conditions in Eq. (\ref{hovering}), the nonlinear UAV Bicopter equation in Eq. (\ref{eq3w44}) can now be written in linear form in Eq. (\ref{linearisasi}).

\begin{equation}
	\begin{split}
		\dot{x}_{1}&=\dot{x}=x_{2}\\
		\dot{x}_{2}&=\ddot{x}=-g\theta \\
		\dot{x}_{3}&=\dot{y}=x_{4}\\
		\dot{x}_{4}&=\ddot{y}=g\phi \\
		\dot{x}_{5}&=\dot{z}=x_{6}\\
		\dot{x}_{6}&=\ddot{z}=g-\frac{u_1}{m}\\
		\dot{x}_{7}&=\dot{\phi} =x_{8}\\
		\dot{x}_{8}&=\ddot{\phi}=\frac{L}{I_{xx}}u_2\\
		\dot{x}_{9}&=\dot{\theta} =x_{10}\\
		\dot{x}_{10}&=\ddot{\theta} =\frac{h}{I_{yy}}u_3\\
		\dot{x}_{11}&=\dot{\psi} =x_{11}\\
		\dot{x}_{12}&=\ddot{\psi} =\frac{L}{I_{zz}}u_4\\
	\end{split}
	\label{linearisasi}
\end{equation}

With the sequence of states described in Eq. (\ref{eq3w32}) - (\ref{eq3w33}), it is possible to rearrange the linear UAV Bicopter model in the form of state space in Eq. (\ref{bicopterlinear}). By knowing the dynamic model parameters of the UAV Bicopter in Table \ref{tab:parameterBicopter}, the values of Matrix A and Matrix B can be obtained for the linear model of the UAV Bicopter.

\begin{figure}[h]
	\centering
	\includegraphics[scale=0.33]{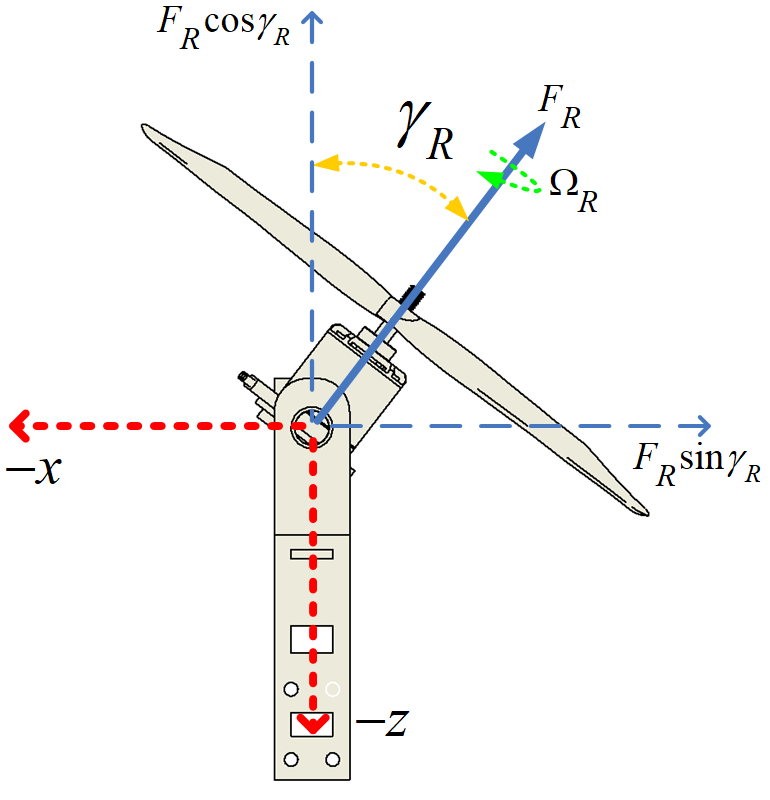}
	\caption{Rotor thrust force due to tilt angle $\gamma  _{R}$ and $\gamma  _{L}$.}
	\label{fig:gb3w15}
\end{figure}

\subsection{Inertial Disturbance Modeling}

By attaching a payload to the UAV Bicopter, it is possible to model inertial disturbances. Previously, it was presumed that the Bicopter operated under nominal conditions ($J_0$), so the inertial matrix was diagonal according to Eq. (\ref{inersianormal}). However, when provided payload, the Bicopter's inertial matrix will undoubtedly change. In order to facilitate the analysis, the inertial matrix is separated into two types: the inertial matrix at nominal conditions and the inertial matrix that results from the presence of payload ($\Delta J$). The value of $\Delta J$ is affected by the products' mass, shape, and position of payload. The addition of the inertial matrix that appears due to the presence of payload referred to as the perturbed inertia matrix, allowing the inertial matrix to be written as Eq. (\ref{inersiatotal}).

\begin{equation}
	J_{0}=\begin{bmatrix}
		I_{xx} & 0 &0 \\ 
		0& I_{yy} &0 \\ 
		0 &  0& I_{zz}
	\end{bmatrix}
	\label{inersianormal}
\end{equation}

\begin{equation}
	J_{\delta}=J_{0}+\Delta J
	\label{inersiatotal}
\end{equation}

\begin{figure}[h]
	\centering
	\includegraphics[scale=0.33]{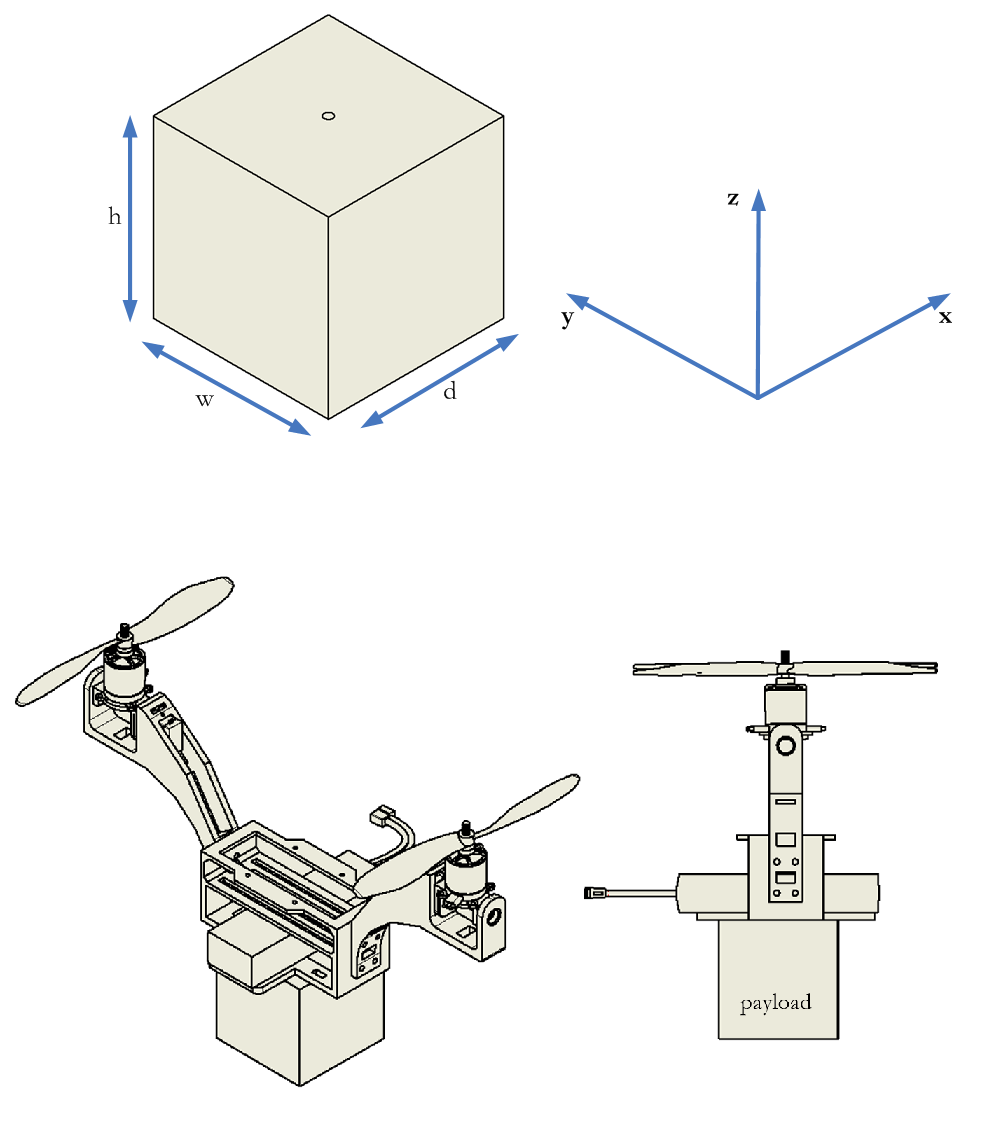}
	\caption{Models of payloads}
	\label{fig:payload}
\end{figure}

It is assumed that the UAV Bicopter's payload will carry a solid cubic-shaped block filled with liquid as shown in Fig. \ref{fig:payload} with the specifications in Table \ref{tab:spesifikasipayload}. Because the payload is in the form of blocks, the inertial disturbance matrix is a symmetric matrix with zero off-diagonal elements as in Eq. (\ref{inersiagangguan}), with the matrix elements consisting in Eq. (\ref{elemengangguan}).

\begin{equation}
	\Delta J=\begin{bmatrix}
		\Delta I_{xx} & 0 &0 \\ 
		0& \Delta I_{yy} &0 \\ 
		0 &  0& \Delta I_{zz}
	\end{bmatrix}
	\label{inersiagangguan}
\end{equation}

\begin{equation}
	\begin{split}
		\Delta I_{xx}=\frac{m}{12}\left ( d^{2}+h^{2} \right )\\
		\Delta I_{yy}=\frac{m}{12}\left ( d^{2}+w^{2} \right )\\
		\Delta I_{zz}=\frac{m}{12}\left ( h^{2}+w^{2} \right )
	\end{split}
	\label{elemengangguan}
\end{equation}

\begin{table}[h]
	\centering
	\caption{Payload specifications.}
	\label{tab:spesifikasipayload}
	\begin{tabular}{p{0.3\linewidth}ll}
		\hline
		Parameter & Value & Units \\ \hline
		Mass     & $0.2$      & $Kg$   \\
		Length ($d$)   & $0.08$    & $m$   \\
		width ($w$)     & $0.08$    & $m$   \\
		Height ($h$)    & $0.08$    & $m$   \\ \hline
	\end{tabular}%
\end{table}

\section{Trajectory Tracking Control of Bicopter using LQG}

\begin{figure}[h]
	\centering
	\includegraphics[scale=0.26]{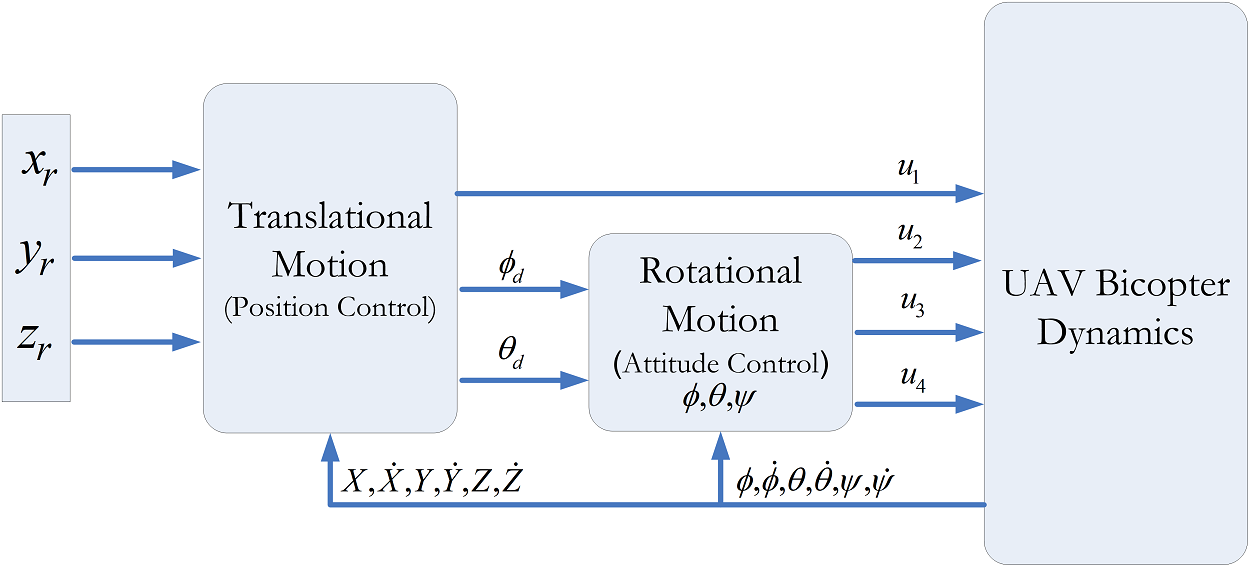}
	\caption{Closed loop control system on UAV Bicopter.}
	\label{fig:gb2w24}
\end{figure}

Figure \ref{fig:gb2w24} shows the block diagram of the closed-loop control system on the UAV Bicopter, which consists of two loops: 1) the inner loop and 2) the outer loop. These two loops are described as follows:

\begin{itemize}
	\item Attitude control (inner loop), controlling the orientation attitude of the UAV in the form of rotational movement. As an inner loop in the UAV system, the rotational motion system must have a fast settling time to support the translational motion system as an outer loop.
	\item Position control (outer loop), controlling the position of the UAV in the form of translational motion. The problem that arises in this position control is tracking. The translational motion system must be able to follow the given reference signal and overcome the given disturbance.
	
\end{itemize}

Using full state feedback control, we can ensure that all eigenvalues of the closed-loop system lie in the left half of the complex plane. This is accomplished by stabilizing a particular system. Consider a linear dynamic system represented by the form  $\dot{x}=Ax+Bu$ in the state space. By utilizing the full state feedback, which represents a linear combination of the state variables, that is $u=-Kx$ in order for the closed-loop system, which is given by Eq. (\ref{feedback}).

\begin{equation}
	\begin{split}
		\dot{x}&=\left ( A-BK \right )x\\
		y&=Cx 
	\end{split}
	\label{feedback}
\end{equation}

\begin{figure*}
	\begin{equation}
		\begin{split}
			A&=\begin{bmatrix}
				0 & 1 & 0 & 0 & 0 & 0 & 0 & 0 & 0 & 0 & 0 & 0\\ 
				0 & 0 & 0 & 0 & 0 & 0 & 0 & 0 & -g & 0 & 0 & 0\\ 
				0 & 0 & 0 & 1 & 0 & 0 & 0 & 0 & 0 & 0 & 0 & 0\\ 
				0 & 0 & 0 & 0 & 0 & 0 & g & 0 & 0 & 0 & 0 & 0\\ 
				0 & 0 & 0 & 0 & 0 & 1 & 0 & 0 & 0 & 0 & 0 & 0\\ 
				0 & 0 & 0 & 0 & 0 & 0 & 0 & 0 & 0 & 0 & 0 & 0\\ 
				0 & 0 & 0 & 0 & 0 & 0 & 0 & 1 & 0 & 0 & 0 & 0\\ 
				0 & 0 & 0 & 0 & 0 & 0 & 0 & 0 & 0 & 0 & 0 & 0\\ 
				0 & 0 & 0 & 0 & 0 & 0 & 0 & 0 & 0 & 1 & 0 & 0\\ 
				0 & 0 & 0 & 0 & 0 & 0 & 0 & 0 & 0 & 0 & 0 & 0\\ 
				0 & 0 & 0 & 0 & 0 & 0 & 0 & 0 & 0 & 0 & 0 & 1\\ 
				0 & 0 & 0 & 0 & 0 & 0 & 0 & 0 & 0 & 0 & 0 & 0
			\end{bmatrix}, \;\;\;\;\;
			B=\begin{bmatrix}
				0 & 0 & 0 & 0\\ 
				0 & 0 & 0 & 0\\ 
				0 & 0 & 0 & 0\\ 
				0 & 0 & 0 & 0\\ 
				0 & 0 & 0 & 0 \\ 
				\frac{1}{m} & 0 & 0 & 0 \\ 
				0 & 0 & 0 &0 \\ 
				0 & \frac{L}{I_{xx}} & 0 & 0\\ 
				0 & 0 & 0 & 0\\ 
				0 & 0 & \frac{h}{I_{yy}} & 0\\ 
				0 & 0 & 0 & 0\\ 
				0 & 0 & 0 & \frac{L}{I_{zz}}
			\end{bmatrix} \\
			C&=\begin{bmatrix}
				1 & 0 & 0 & 0 & 0 & 0 & 0 & 0 & 0 & 0 & 0 & 0\\ 
				0 & 1 & 0 & 0 & 0 & 0 & 0 & 0 & 0 & 0 & 0 & 0\\ 
				0 & 0 & 1 & 0 & 0 & 0 & 0 & 0 & 0 & 0 & 0 & 0\\ 
				0 & 0 & 0 & 1 & 0 & 0 & 0 & 0 & 0 & 0 & 0 & 0\\ 
				0 & 0 & 0 & 0 & 1 & 0 & 0 & 0 & 0 & 0 & 0 & 0\\ 
				0 & 0 & 0 & 0 & 0 & 1 & 0 & 0 & 0 & 0 & 0 & 0\\ 
				0 & 0 & 0 & 0 & 0 & 0 & 1 & 0 & 0 & 0 & 0 & 0\\ 
				0 & 0 & 0 & 0 & 0 & 0 & 0 & 1 & 0 & 0 & 0 & 0\\ 
				0 & 0 & 0 & 0 & 0 & 0 & 0 & 0 & 1 & 0 & 0 & 0\\ 
				0 & 0 & 0 & 0 & 0 & 0 & 0 & 0 & 0 & 1 & 0 & 0\\ 
				0 & 0 & 0 & 0 & 0 & 0 & 0 & 0 & 0 & 0 & 1 & 0\\ 
				0 & 0 & 0 & 0 & 0 & 0 & 0 & 0 & 0 & 0 & 0 & 1
			\end{bmatrix},\;\;\;\;\;
			D=\begin{bmatrix}
				0 & 0 & 0 & 0\\ 
				0 & 0 & 0 & 0\\ 
				0 & 0 & 0 & 0\\ 
				0 & 0 & 0 & 0\\ 
				0 & 0 & 0 & 0\\ 
				0 & 0 & 0 & 0\\ 
				0 & 0 & 0 & 0\\ 
				0 & 0 & 0 & 0\\ 
				0 & 0 & 0 & 0\\ 
				0 & 0 & 0 & 0\\ 
				0 & 0 & 0 & 0\\ 
				0 & 0 & 0 & 0
			\end{bmatrix}
		\end{split}
		\label{bicopterlinear}
	\end{equation}
\end{figure*}

One type of full-state feedback optimal control is the linear quadratic regulator (LQR). The optimality criterion for LQR is specified by the cost function in Eq. (\ref{JLQR}). The cost of each state $x$ and control input $u$ for a system defined in linear state space is represented by matrices $Q$ and $R$. Calculating the control inputs that will yield the lowest possible value of cost function $J$ is referenced in Eq. (\ref{uLQR}). The continuous Ricatti equation has a solution of $P$, as in Eq. (\ref{ricati}).

\begin{equation}
	J=\int_{0}^{\infty }\left [ x_{(t)}^{T}Qx_{(t)} +u_{(t)}^{T}Ru_{(t)}\right ]dt
	\label{JLQR}
\end{equation}

\begin{equation}
	u=-Kx=-R^{-1}B^{T}Px
	\label{uLQR}
\end{equation}

\begin{equation}
	A^{T}P+PA+Q-PBR^{-1}B^{T}P=0
	\label{ricati}
\end{equation}

\begin{figure*}
	\centering
	\includegraphics[scale=0.3]{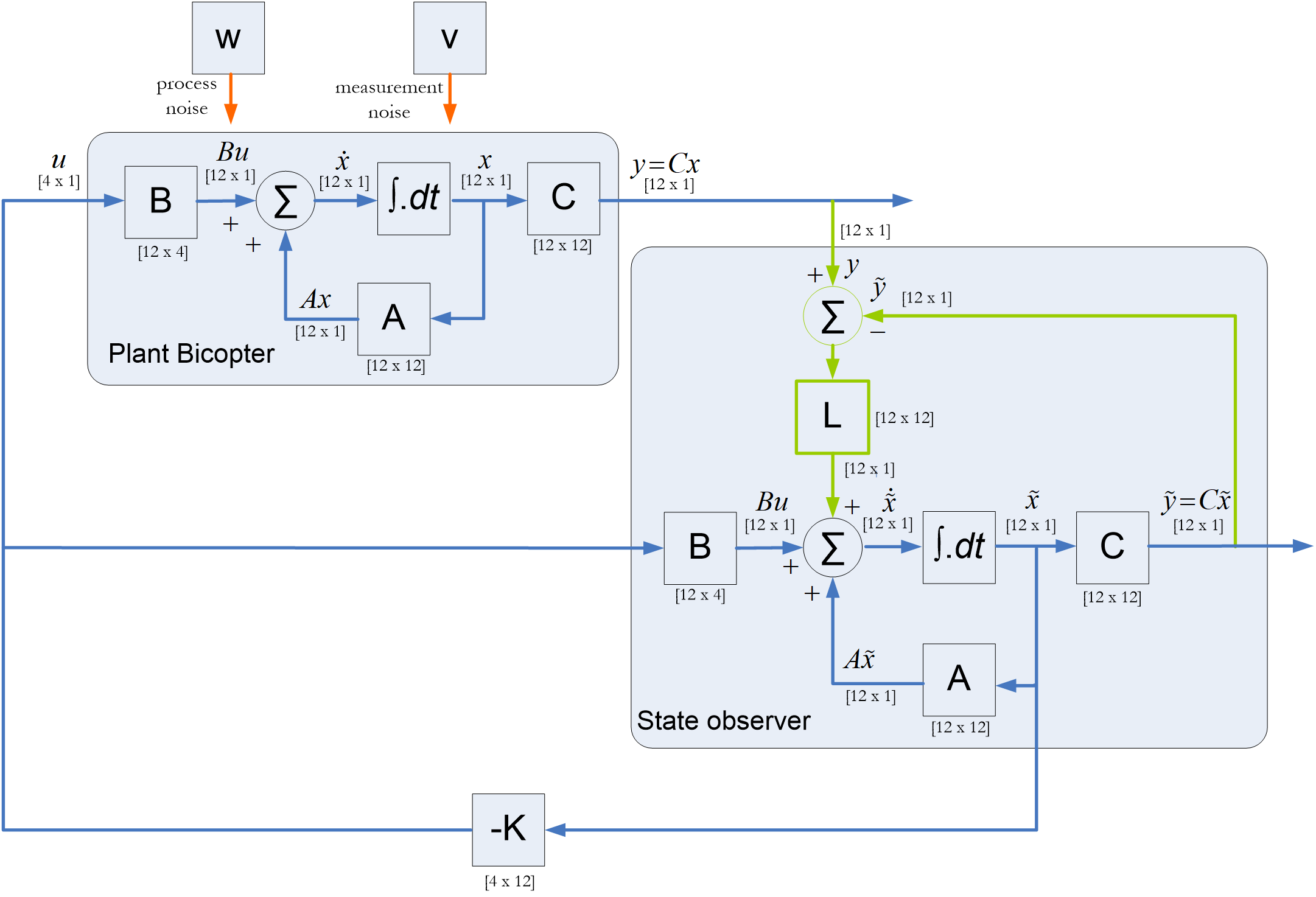}
	\caption{LQG controller design on UAV Bicopter.}
	\label{fig:gb2w35}
\end{figure*}

The use of LQR control will be optimal in the case of a state regulator towards an equilibrium point in a system, but for the tracking case, the LQR controller is not capable of optimally, especially in systems that have high levels of nonlinearity and system dynamic changes such as inertial disturbances. Therefore, this paper proposes the design of LQG control, which is expected to overcome system dynamic changes, in this case, in the form of inertial disturbances to the Bicopter when carrying a payload. The LQG controller design that will be applied to the Bicopter flight simulation is described in Fig. \ref{fig:gb2w35} as follows:

\begin{itemize}
	\item Requires a Bicopter dynamics model in the form of state space
	\item Doing the process of linearization of the model
	\item Check the controlability and observability of the system
	\item Adding the noise process covariance Matrix, $w$ and noise measurement covariance, $v$
	\item Determine the $Q$ Matrix and $R$ Matrix
	\item Calculate the feedback gain $L$ 
	\item Calculate the feedback gain $K$
	\item After the feedback gain $K$ is obtained, then solve the differential equation of the Bicopter model with "ode45".	
\end{itemize}

Figure \ref{fig:gb2w35} presents the Bicopter state space in Eq. (\ref{eq2w56}) and the state observer in Eq. (\ref{eq2w57}). In the LQG controller, there is an observer whose purpose is to determine the estimate of $\tilde{x}$ so that the estimated value approaches the actual value  $\tilde{x}\rightarrow x$ along with $t\rightarrow \infty$. Remember that the value of $x(t_{0})$ is uncertain. Calculating the initial value of $\tilde{x}(t_{0})$ to the observer is necessary by defining the estimation error as in Eq. (\ref{eq2w58}).

\begin{equation}
	\dot{x}=Ax+Bu; ~y=Cx
	\label{eq2w56}
\end{equation}

\begin{equation}
	\dot{\tilde{x}}=A\tilde{x}+Bu+L\left ( y-\tilde{y} \right ); ~\tilde{y}=C\tilde{x}
	\label{eq2w57}
\end{equation}

\begin{equation}
	e(t)=x(t)-\tilde{x}(t)
	\label{eq2w58}
\end{equation}

The observer is designed with the property  $e(t)=x(t)-\tilde{x}(t)$ along with $t\rightarrow \infty $. If the system has fully observable properties, then the $L$ Matrix will always be obtained so the tracking error is asymptotically stable as desired. By deriving Eq. (\ref{eq2w58}) against time, Eq. (\ref{eq2w59}) is obtained.

\begin{equation}
	\begin{split}
		\frac{d}{dt}e(t)&=\left ( \frac{d}{dt} x(t)\right )-\left ( \frac{d}{dt}\tilde{x}(t) \right ) \\
		&=\left ( Ax+Bu \right )-\left ( A\tilde{x}+Bu+L\left ( y-\tilde{y} \right ) \right ) \\
		&=\left ( Ax+Bu \right )-\left ( A\tilde{x}+Bu+LCx-LC\tilde{x} \right ) \\
		&=A\left ( x-\tilde{x} \right )+LC\left ( \tilde{x}-x \right )\\
		&=A\left ( x-\tilde{x} \right )-LC\left ( x-\tilde{x} \right )\\
		&=\left ( A-LC \right )\left ( x-\tilde{x} \right ) \\
		&=\left ( A-LC \right )e \\
	\end{split}
	\label{eq2w59}
\end{equation}

Equation (\ref{eq2w59}) is an estimate error dynamics system, with $e(t)\rightarrow 0$ as $t\rightarrow \infty$ for each initial tracking error $e(t_{0})$ guaranteed by the characteristic equation in Eq. (\ref{eq2w60}) by having all roots on the left side of the imaginary axis in the plane $s$. Therefore, the observer design process determines the $L$ Matrix so that the roots of the characteristic equation are on the left side of the imaginary plane. This is achieved if the system has fully observable.

\begin{equation}
	\left | \lambda I-\left ( A-LC \right ) \right |=0
	\label{eq2w60}
\end{equation}

After the $w$ and $v$ matrices have been determined, the gain $L$, also known as the Kalman gain can be acquired by executing the command $"L = lqe(A, w, C, w, v)"$ in MATLAB. When the Bicopter has noise in the form of process noise ($w$) and measurement noise ($v$) as in Eq. (\ref{sistemnoise}) so the estimated error dynamics system becomes Eq. (\ref{errordynamic}).

\begin{equation}
	\begin{split}
		\dot{x}&=Ax+Bu+w\\
		y&=Cx+v\\
	\end{split}
	\label{sistemnoise}
\end{equation}

\begin{equation}
	\begin{split}
		\dot{e}&=Ax+Bu+w-A\tilde{x}-Bu-L\left ( Cx+v-C\tilde{x} \right )\\
		&=\left ( A-LC \right )e+w-Lv
	\end{split}
	\label{errordynamic}
\end{equation}

In Eq. (\ref{errordynamic}), there exist $w$ and $v$, which means that the estimation is not going to be zero in most cases, so if we want the error to remain small we can do it with the right choice $L$, the optimal choice $L$ is given by the Kalman gain.

Once the linear model of the UAV Bicopter has been known and the $Q$ Matrix and $R$ Matrix have been identified, the $w$ Matrix and $v$ Matrix have been decided, and the feedback gain matrix $L$ has been obtained, the feedback gain $K$ may be computed using the LQR technique. We are able to determine the feedback gain $K$ with the assistance of MATLAB by utilizing the command $"K = lqr((A-LC),B,Q,R)"$.

The results of the feedback gain $K$ obtained from the LQG controller will be used as input control, $u=-Kx$. Then, use the "ode45" function in MATLAB to solve the differential equation of the UAV Bicopter in Eq. (\ref{eq3w44}).

\newpage
\section{Simulation Results and Discussion}
To validate the effectiveness of the proposed control, a simulation is presented in MATLAB. Table \ref{tab:parameterBicopter} shows the Bicopter parameters for simulations. The Bicopter had to follow the desired circular trajectory defined in (\ref{melingkar}) and trajectory number "8" in (\ref{angka8}).

\begin{equation}
	\begin{split}
		x_{d}&=-1,3~cos\left ( \pi t  \right )\\
		y_{d}&=1,3~sin\left ( \pi t  \right )\\
		z_{d}&=2 \\
		\psi_d&=0
	\end{split}
	\label{melingkar}
\end{equation}

\begin{equation}
	\begin{split}
		x_{d}&=sin\left ( 0,25\pi t  \right )\\
		y_{d}&=sin\left (  0,5\pi t  \right )\\
		z_{d}&=2 \\
		\psi_d&=0
	\end{split}
	\label{angka8}
\end{equation}

\begin{table}[h]
	\centering
	\caption{Bicopter dynamic model parameters.}
	\label{tab:parameterBicopter}
	\resizebox{\columnwidth}{!}{%
		\begin{tabular}{p{0.45\linewidth}ccl}
			\hline
			Parameter                                               & Symbols & Value  & \multicolumn{1}{c}{Unit} \\ \hline
			Mass of the UAV Bicopter                                & $m$       & 0.725   & $kg$                       \\
			Gravitational acceleration                              & $g$       & 9.81   & $m.s^{-2}$                     \\
			Vertical distance between CoG and center   of the rotor & $h$       & 0.042  & $m$                        \\
			Horizontal distance CoG and rotor   center              & $L$       & 0.225  & $m$                        \\
			Thrust coefficient                                      & $C_T$      & 0.1222 &   -                       \\
			The Moment of Inertia along x axis                      & $I_{xx}$     & $0.116 \times10^{-3}$  & $kg.m^{2}$                    \\
			The Moment of Inertia along y axis                      & $I_{yy}$     & $0.0408 \times10^{-3}$ & $kg.m^{2}$                    \\
			The Moment of Inertia along z axis                      & $I_{zz}$     & $0.105 \times10^{-3}$  & $kg.m^{2}$                    \\ \hline
		\end{tabular}%
	}
\end{table}

\begin{figure}[h]
	\centering
	\includegraphics[scale=0.6]{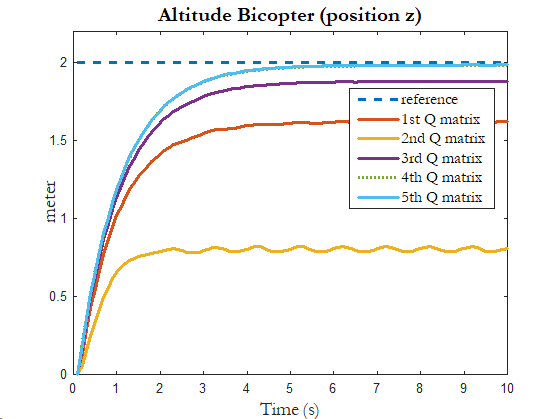}
	\caption{Altitude response when heading to position $z = 2$ with variations in the $Q$ Matrix weighting.}
	\label{fig:altituderespons}
\end{figure}

In the first test scenario, the altitude is the only information given about the desired trajectory. Two meters was chosen to serve as our point of reference for altitude. When using the LQR controller, the $Q$ Matrix and $R$ Matrix parameters will significantly affect the feedback gain $K$ results. In the early stages of determining the parameters of the $Q$ and $R$ matrices, a process of tuning the $Q$ and $R$ weights was carried out with the values $Q=[C^TC]$ and $R=[eye(4)\times 0,001]$. Furthermore, by changing the $Q$ weighting matrix, the feedback gain $K$ variation will be obtained. As indicated in Table \ref{tab:variasiQ}, this research employed a $Q$ weighting matrix with a range of five variations to assess the value of variation. Fig. \ref{fig:altituderespons} presents the altitude response of the UAV Bicopter with variations in the $Q$ Matrix weighting. This altitude response graph shows that the 5th $Q$ Matrix weighting parameter produces the smallest root mean square error (RMSE) compared to the others and does not experience overshoot as shown in Table \ref{tab:karakteristik altitude}.

\begin{table}[h]
	\centering
	\caption{Variation of weighting of the Q Matrix.}
	\label{tab:variasiQ}
	\begin{tabular}{cl}
		\hline
		\multicolumn{1}{l}{{\scriptsize No}} & {\scriptsize Q weighting matrix} \\ \hline
		 1                      &   $Q=[C^TC]$                   \\
		 2                      &   $Q=[eye\left ( 12 \right )\times 0,1]$                  \\
		 3                      &   $Q=[eye\left ( 12 \right )\times 10]$                  \\
		 4                     &   $Q=[eye\left ( 12 \right )\times 300]$                  \\
		 5                      &  $Q=[eye\left ( 12 \right )\times 700]$                  \\ \hline
	\end{tabular}%
\end{table}

\begin{table}[h]
	\centering
	\caption{Response characteristics at altitude Bicopter with LQR control.}
	\label{tab:karakteristik altitude}
	\resizebox{\columnwidth}{!}{%
		\begin{tabular}{lccccc}
			\hline
			\multicolumn{1}{c}{\multirow{2}{*}{Characteristics}} & \multicolumn{5}{c}{Parameters of the Q weighting matrix} \\ \cline{2-6} 
			\multicolumn{1}{c}{} & 1      & 2      & 3      & 4      & 5      \\ \hline
			RiseTime (s)            & 2.0138 & 1.0619 & 2.1381 & 2.1696 & 2.1717 \\
			SettlingTime (s)        & 3.8157 & 9.7043 & 3.8557 & 3.8689 & 3.8699 \\
			SettlingMin          & 1.4725 & 0.7311 & 1.7024 & 1.7855 & 1.7919 \\
			SettlingMax          & 1.6258 & 0.8206 & 1.8816 & 1.9782 & 1.9857 \\
			Overshoot (\%)             & 0.1095 & 1.3647 & 0.0227 & 0      & 0      \\
			RMSE                 & 0.6339 & 1.2574 & 0.5016 & 0.4722 & 0.4705 \\ \hline
		\end{tabular}%
	}
\end{table}

Furthermore, testing was carried out by adding inertial disturbances to the Bicopter UAV dynamics system. This inertial disturbance is in the form of a payload that will be carried by the Bicopter in following a given trajectory. Based on Eq. (\ref{elemengangguan}) and Table \ref{tab:spesifikasipayload} related to the mass and dimensions of the payload to be carried are described in Table \ref{tab:inersia}.

\begin{table}[H]
	\centering
	\caption{Inertia of the UAV Bicopter after being given a payload.}
	\label{tab:inersia}
	\resizebox{\columnwidth}{!}{%
		\begin{tabular}{lcccl}
			\hline
			Parameter & Initial inertia & Payload inertia & Total    & \multicolumn{1}{c}{Units} \\ \hline
			$I_{xx}$       & 1.16E-04     & 2.13E-04        & 3.29E-04 &   $kg.m^{2}$          \\
			$I_{yy}$       & 4.08E-05     & 2.13E-04        & 2.54E-04 &     $kg.m^{2}$        \\
			$I_{zz}$      & 1.05E-04     & 2.13E-04        & 3.18E-04 &       $kg.m^{2}$       \\ \hline
		\end{tabular}%
	}
\end{table}

\begin{figure}[H]
	\begin{subfigure}{.5\textwidth}
		\centering
		\includegraphics[scale=0.29]{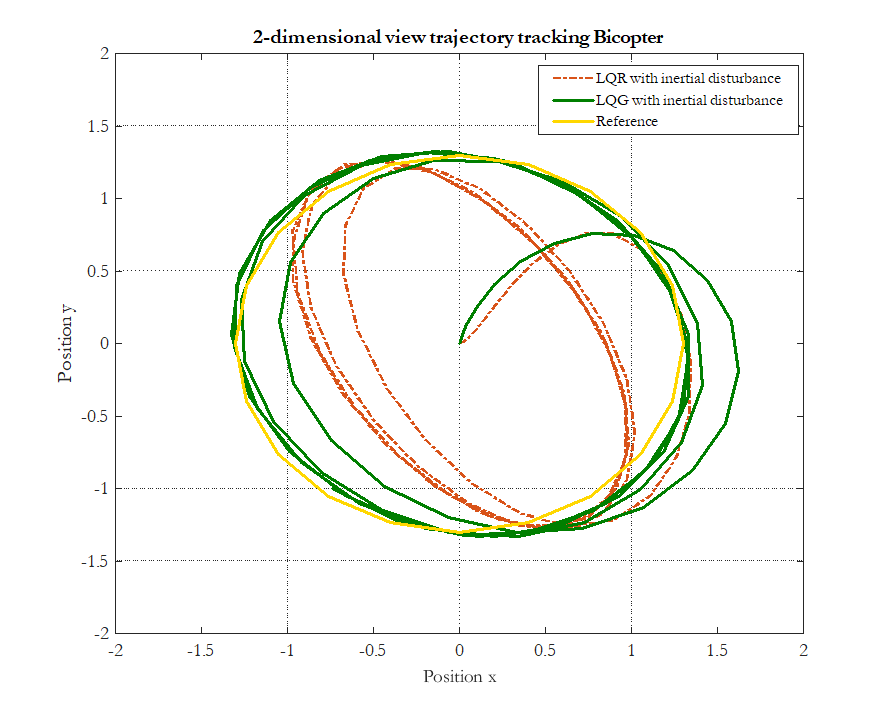}
		\caption{2D view circular trajectory}
	\end{subfigure}
	\begin{subfigure}{.5\textwidth}
		\centering
		\includegraphics[scale=0.37]{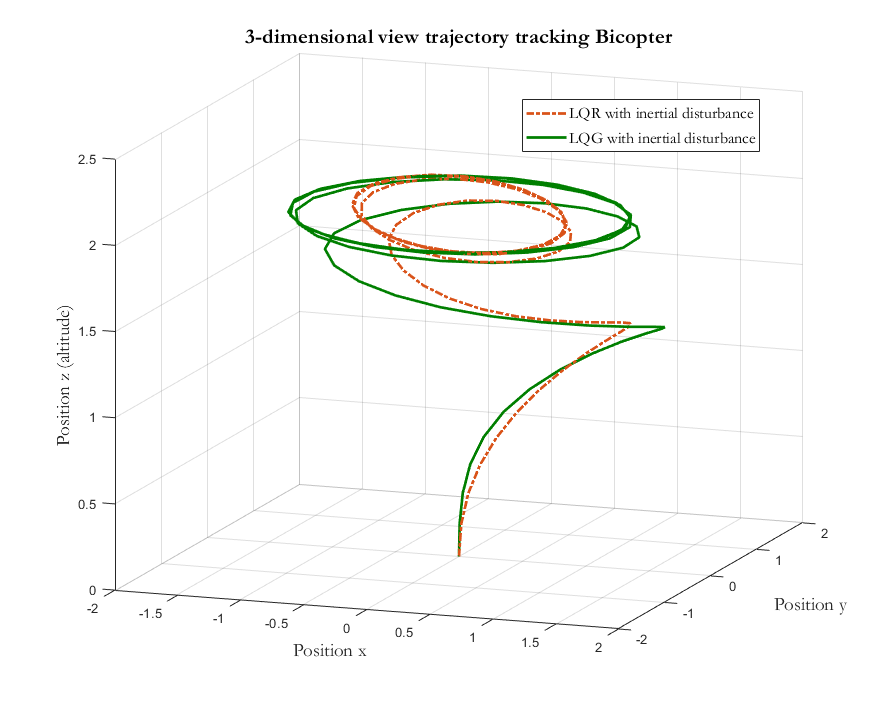}
		\caption{3D view circular trajectory}
	\end{subfigure}
	\caption{Bicopter position response using LQR vs. LQG during circle trajectory tracking with inertial disturbance, the video recording of the test results can be seen in the following link: https://youtu.be/R5Ak8hxK9GM .}
	\label{lingkaran tracking}
\end{figure}

\begin{figure}[H]
	\begin{subfigure}{.5\textwidth}
		\centering
		\includegraphics[scale=0.29]{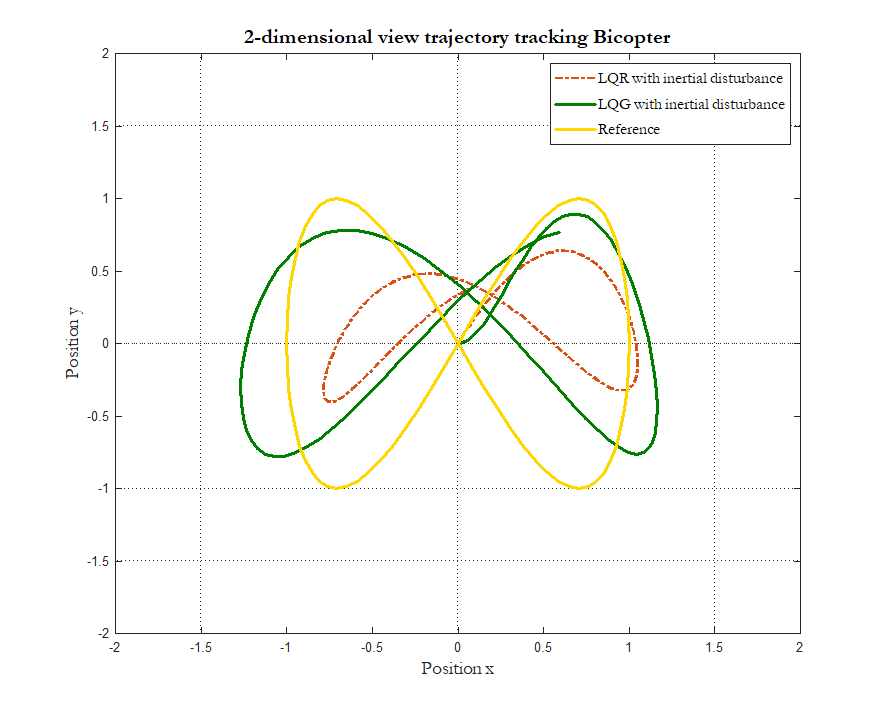}
		\caption{2D view “8” trajectory}
	\end{subfigure}
	\begin{subfigure}{.5\textwidth}
		\centering
		\includegraphics[scale=0.37]{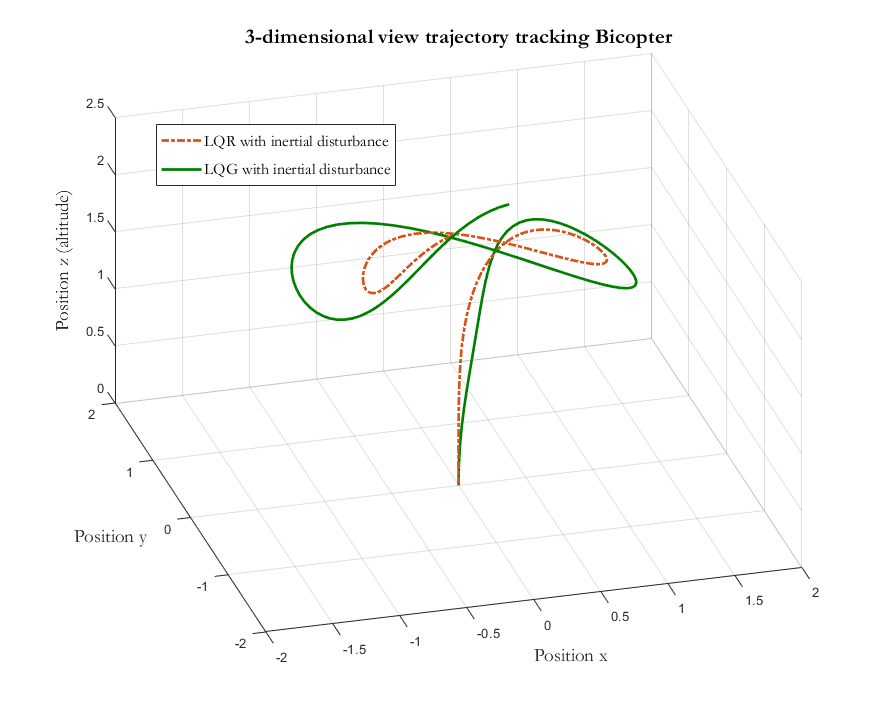}
		\caption{3D view “8” trajectory}
	\end{subfigure}
	\caption{Bicopter position response using LQR vs. LQG during "8" trajectory tracking with inertial disturbance, the video recording of the test results can be seen in the following link: https://youtu.be/stBaKI2x1qU .}
	\label{8 tracking}
\end{figure}

By using the LQG controller, the addition of the $w$ Matrix and $v$ Matrix is set with the value $w = [eye(12) \times 0,01]$ and Matrix  $v = [eye(12) \times 10]$. Figure \ref{lingkaran tracking} shows the results of the response after being given an inertial disturbance for circle trajectory tracking and the trajectory number ”8” in Fig. \ref{8 tracking} when given an inertial disturbance. Table \ref{RMSElingkaran} presents the RMSE value when the UAV Bicopter follows a circular trajectory and Table \ref{RMSE8} when it follows the “8” trajectory.

\begin{table}[h]
	\centering
	\caption{RMSE value when following a circular trajectory tracking using LQR vs LQG control with inertial disturbance.}
	\label{RMSElingkaran}
	\begin{tabular}{lcc}
		\hline
		Position   & LQR  & LQG \\ \hline
		 Position $x$ & 1.0204                 &  0.8630                 \\
		 Position $y$ &  0.7453                 &  0.7426                \\ \hline
	\end{tabular}%
\end{table}

From Table \ref{RMSElingkaran}, it can be observed that when there is inertia disturbance, the LQG controller can significantly reduce RMSE at position $x$ and slightly decrease RMSE at position $y$. Likewise, when following the trajectory of the number "8". Table \ref{RMSE8} shows that the RMSE value of the $x$ and $y$ positions also decreases when using the LQG control. This means that the LQG control method is able to track the desired trajectory more accurately than the LQR control method. The reason why the LQG control method has better tracking performance is that it takes into account the uncertainty in the Bicopter's dynamics and the disturbances in the environment. The LQR control method does not take into account these uncertainties, which is why it has worse tracking performance.

\begin{table}[h]
	\centering
	\caption{RMSE value when following a "8" trajectory tracking using LQR vs LQG control with inertial disturbance.}
	\label{RMSE8}
	\begin{tabular}{lcc}
		\hline
		Position   & LQR  & LQG \\ \hline
		 Position $x$ &  1.1891                & 0.7395                  \\
		 Position $y$ &  0.9146               &  0.9037                \\ \hline
	\end{tabular}%
\end{table}

\section{Conclusion}
The design of the trajectory tracking controller on the UAV Bicopter was developed using a linear quadratic gaussian (LQG) controller and has been successfully simulated using MATLAB. The simulation results of the UAV Bicopter trajectory tracking using the LQG controller have suppressed the effect of inertial interference in the form of payload in the trajectory tracking case on the UAV Bicopter. The success of LQG control was tested in two scenarios, the first trajectory tracking in the form of a circular position and the second in the form of a trajectory tracking number "8". The simulation results show that the proposed LQG controller overcomes inertial disturbances when the UAV Bicopter performs trajectory tracking, and the results show that the LQG controller has a smaller root mean square error (RMSE) value when compared to using the LQR control.

\section*{Acknowledgment}
This work was supported by the Indonesian Postgraduate Domestic Education Scholarship (BPPDN) with contract number 2974/UN1.P.IV/KPT/DSDM/2019.

\bibliographystyle{IEEEtran}
\bibliography{ref}

\begin{IEEEbiography}[{\includegraphics[width=1in,height=1.25in,clip,keepaspectratio]{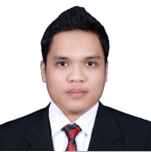}}]{Fahmizal} 
	received the B. Eng. degree in electrical engineering from Institut Teknologi Sepuluh Nopember (ITS), Indonesia in 2012. He graduated from Master of Science in Taiwan, precisely at National Taiwan University of Science and Technology (Taiwan Tech) in 2014. Start from 2015, he is a junior lecturer in the Department of Electrical and Informatics Engineering at Vocational College Universitas Gadjah Mada, Indonesia, in the specialist field of the control system and robotics.  Currently,  he is working toward the doctoral degree in  Department  of  Electrical  and  Information  Engineering,  Engineering  Faculty, Universitas  Gadjah  Mada,  Yogyakarta,  Indonesia. E-mail: fahmizal@ugm.ac.id
\end{IEEEbiography}

\begin{IEEEbiography}[{\includegraphics[width=1in,height=1.25in,clip,keepaspectratio]{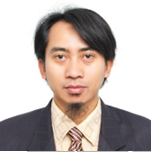}}]{Hanung Adi Nugroho} 
	received the B. Eng. degree in electrical engineering from Universitas Gadjah Mada, Indonesia, 2001 and Master of Engineering degree in Biomedical Engineering from the University of Queensland, Australia in 2005. In 2012, he received his Ph.D. degree in Electrical and Electronic Engineering from Universiti Teknologi Petronas, Malaysia. Currently, he is a Professor and also a Head of Department of Electrical Engineering and Information Technology, Faculty of Engineering, Universitas Gadjah Mada, Indonesia. His current research interests include biomedical signal and image processing and analysis, computer vision, medical instrumentation and pattern recognition. E-mail: adinugroho@ugm.ac.id 
\end{IEEEbiography}

\begin{IEEEbiography}[{\includegraphics[width=1in,height=1.25in,clip,keepaspectratio]{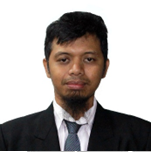}}]{Adha Imam Cahyadi} 
	received the B. Eng. degree in electrical engineering from University of Gadjah Mada, Indonesia in 2002. Then he worked as an engineerin industry, such as in Matsushita Kotobuki Electronics and Halliburton Energy Services for a year. He received the M. Eng. degree in control engineering from King Mongkuts Institute of Technology Ladkrabang, Thailand (KMITL) in 2005, and received the D. Eng. degree in control engineering from Tokai University, Japan in 2008. Currently, he is an Associate Professor at Department of Electrical Engineering and Information Technology, Faculty of Engineering, Universitas Gadjah Mada, Indonesia and a visiting lecturer at the Centre for Artificial Intelligence and Robotics (CAIRO), University of Teknologi Malaysia, Malaysia. His research interests include teleoperation systems and robust control for delayed systems especially process plant. E-mail: adha.imam@ugm.ac.id
\end{IEEEbiography}

\begin{IEEEbiography}[{\includegraphics[width=1in,height=1.25in,clip,keepaspectratio]{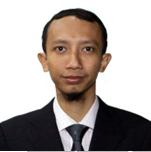}}]{Igi Ardiyanto} 
	received the B. Eng. degree in electrical engineering from University of Gadjah Mada, Indonesia in 2009. The M.Eng. and D. Eng. degrees in computer science and engineering from Toyohashi University of Technology (TUT), Japan in 2012 and 2015, respectively. He joined the TUT-NEDO (New Energy and Industrial Technology Development Organization, Japan) research collaboration on service robots, in 2011. He is now an Associate Professor at Department of Electrical Engineering and Information Technology, Faculty of Engineering, Universitas Gadjah Mada, Indonesia. He received several awards, including Finalist of the Best Service Robotics Paper Award at the 2013 IEEE International Conference on Robotics and Automation (ICRA 2013) and Panasonic Award for the 2012 RT-Middleware Contest. His research interests include planning and control system for mobile robotics, deep learning, and computer vision. E-mail: igi@ugm.ac.id
\end{IEEEbiography}

\EOD

\end{document}